\documentclass{jetpl}
\usepackage{cite}
\twocolumn

\lat

\title{Dynamics of particles trapped by dissipative solitons}

\rtitle{Dynamics of particles trapped by \ldots}

\sodtitle{Dynamics of particles trapped by dissipative solitons}

\author{D.\,A.\,Dolinina, A.\,S.\,Shalin, A.\,V.\,Yulin\thanks{e-mail: alex.v.yulin@gmail.com}}

\rauthor{Dolinina D.A., Shalin A.S., Yulin A.V.}

\sodauthor{ Dolinina, Shalin, Yulin,}

\address{ITMO University,
197101 Saint-Petersburg, Russia}

\dates{}{}

\abstract{Optomechanical manipulation of nanoparticles enabling ultimate control over their 3D motion is nowadays one of the most highly demanded links between optics, biology, medicine, microfluidics, etc., paving the way for a plethora of emerging applications from drug delivery to living cells, to new methods of nanofabrication. In this Letter we provide novel type of optical manipulation driven by nonlinear effects and laying on the interface between classical optomechanics and non-linear optics. The formation, stability and the dynamics of optical dissipative solitary waves interacting with dielectric nanoparticles are studied theoretically. A mathematical model describing the optical field and the particles are proposed and the stationary solutions in the form of localized optical waves interacting with nanoparticles are found, their bifurctations are studied. It is shown that the linear stability of the solitary waves is affected by the particles but there are regions in the parameter space where the solitons remain stable. The dynamics of the solitary waves with trapped nanoparticles under the action of the inhomogeneous pump is also studied.}


\begin{document}

\maketitle

\section{INTRODUCTION}

Optical trapping \cite{Ashkin1, Ivinskaya1, Kostina} and transporting \cite{Ruffner, Petrov, Shalin, Sukhov, Ivinskaya2} of small objects by optical forces has been actively developing in the past several decades. Such optical transport can be widely used in many areas from optics \cite{Moffitt}, biological \cite{Fazal} and chemical \cite{Li} research, to microfabrication \cite{Agarwal}.

Optical trapping is based on the balance of two different types of optical forces. The first type is scattering forces, which push an object in the direction of light propagation; and the second type is gradient optical forces, which move an object along the gradient of light intensity \cite{Ashkin2}. If gradient forces are larger than scattering, then an object is pulled to the area of stronger light intensity and can be trapped by focused light. After trapping an object can be manipulated by moving light beam in a preferred direction.

\begin{figure}[htbp]
\centering
\fbox{\includegraphics[width=\linewidth]{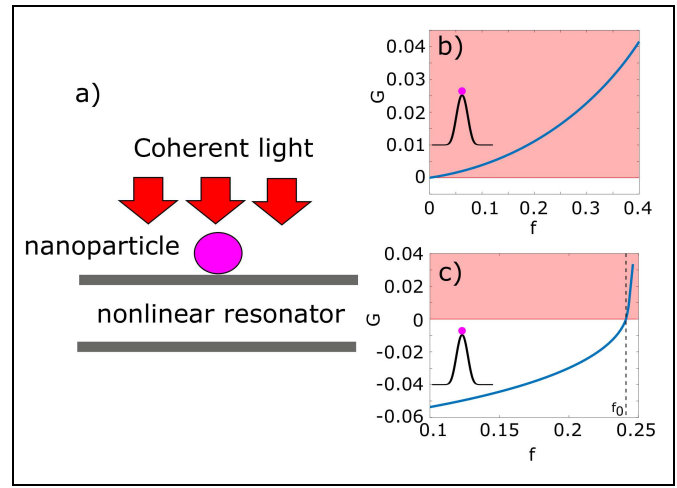}}
\caption{Fig.1 (a) Schematic view of a Fabry-Perot resonator with a nanoparticle on the upper translucent mirror; (b) Instability increment of the resting soliton with a particle placed in the center of soliton; the pumping field is uniform. Parameters: $\nu = 0.7$, $E_0 = 1$, $\delta = 0.3$, $\gamma = 0.2$, $C = 16$, $\alpha = 3$; (c) Same as in (b), but pumping field has phase dependence on coordinate in the form $P = P_0 e^{-i k x^2}$. $k = 0.005$, other parameters the same as in (b).}
\label{fig1}
\end{figure}

In this letter we propose a new concept of optical manipulation of a small particle by dissipative optical localized waves opening a room of opportunities for flexible control over nanoparticles dynamics governed by nonlinear effects. Further we refer the localized waves as solitons. First, the formation of the solitons can locally enhance the intensity of the field and thus allow to use less powerful holding beam. Secondly, changing the phase gradient of the holding beam, (which is a relatively easy task)  it is possible to move the solitons in a controllable way and if the particles are trapped by soliton, then the particles will be moved by it. This mechanism can potentially facilitate the manipulation of nanoobjects by light.

\begin{figure}[t]
\centering
\fbox{\includegraphics[width=\linewidth]{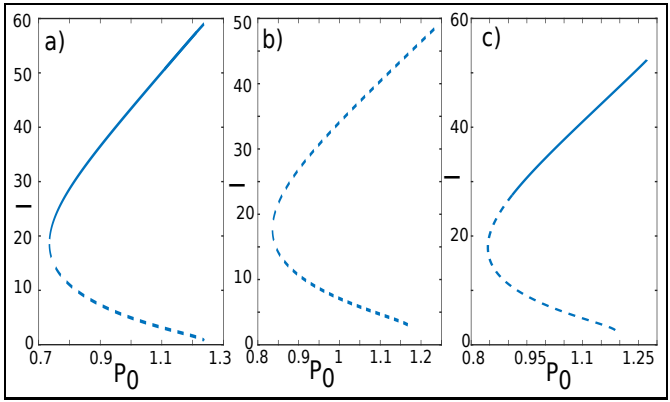}}
\caption{Fig.2 (a) Bifurcation curve showing the dependence of peak soliton intensity on pumping in case of the absence of a particle and uniform pumping $P = P_0$. Dashed line shows dynamically unstable solutions. $f = 0.2$ for all panels; (b) Bifurcation curve for the resting soliton with a trapped particle under uniform pumping. All solutions are unstable; (c) Bifurcation curve for the resting soliton with a trapped particle under pumping of the form $P = P_0 e^{-ikx^2}$, $k = 0.0005$. The upper soliton branch is partly unstable.}
\label{fig4}
\end{figure}

To implement this strategy of optical manipulation the number of issues has to be clarified. First, it is necessary to suggest a mathematical model describing the formation of the solitons in the presence of nanoparticles. Secondly, stationary localized solutions with trapped particles have to be found and the effect of the particles on the stability of the  solitons must be addressed too. Finally, the trapped particles can affect the motion of the solitons significantly, and this effect has to be studied by direct numerical simulations.

\section{CONSIDERED MODEL}

The physical system considered in the present paper is a nonlinear Fabry-Perot resonator with dielectric nanoparticles placed on top of the resonator. The resonator is pumped by a holding beam of coherent light. The schematic view of this system is shown in fig.\ref{fig1}(a).

Bistability and formation of the solitons in such resonators have been studied for many years theoretically \cite{Szoke, Firth, Rosanov} and experimentally \cite{Gibbs,Firth2}. The novelty of the present Letter is in the presence of the nanoparticles that partially screen the light exciting the resonator. At the same time spatially nonuniform distribution of the optical field in the resonator leads to the lateral force acting on the particle. In the simplest case this force is proportional to the optical field  intensity gradient, and this ponderomotive force drags a particle into an area with higher intensity. The particles are small, so it's possible to neglect their inertia and consider their motion as a viscous one. In turn, the nanoparticle creates a shadow which locally reduces the intensity of the coherent pumping of the optical mode. Hence optical field affects particle's dynamics, and particle's location, in turn, influences optical mode pumping.

As a mathematical model describing the dynamics of light in such system, the generalized nonlinear Schrödinger equation with dissipation and pumping can be used. The self-consistent system of equations is:

\begin{eqnarray}
    \dfrac{\partial}{\partial t} E - i C \dfrac{\partial^2}{\partial x^2} E + (\gamma - i \delta - i \dfrac{\alpha}{1 + |E|^2})E = \nonumber \\
    = (1 - f e^{-(x - \epsilon)^2/\omega^2})P, \\
    \label{eq:schrod}
    \dfrac{\partial}{\partial t} \epsilon = \eta \dfrac{\partial}{\partial x}  \alpha |E(\epsilon)|^2,
    \label{eq:part}
\end{eqnarray}
where $E$ - is a complex amplitude of optical field in the resonator, $P$ - amplitude of laser pumping, $\gamma$ - decay rate, $\alpha$ - coefficient of nonlinearity; $\delta$ - laser detuning from resonant frequency, $\epsilon$ - coordinate of nanoparticle. Parameter $\omega$ defines width of a particle shadow, $f$ defines transparency of a particle: if $f = 0$, then the particle is transparent and if $f=1$ then the particle is opaque. The coefficient $\eta$ defines the ratio of the dragging force acting on the particle to the  field intensity gradient in the point of the particle location. Let us note that for mathematical convenience we use the dimensionless variables.

We studied numerically how the presence of particles affects the stability and dynamics of solitons. It was found that the destruction of an optical dissipative soliton occurs only if the presence of a particle leads to a sufficiently strong local decrease in pumping. Also, the dynamics of solitons can significantly change under the particle's influence. In next Sections we show how stability of a soliton changes with the presence of a particle and consider it's dynamics.

\section{RESTING SOLITON WITH CAPTURED PARTICLE}

It is obvious that a particle is resting only if the spatial derivative of the optical field intensity is equal to zero. If the resting soliton has only one maximum (one-hump soliton), then the resting particle can be either in the centre of the soliton or infinitely from it. The latter case is trivial and we focus on more interesting case when the particle is situated in the soliton centre.

Let us briefly describe the bifurcation diagram of the solitons under uniform pumping $P = P_0$ and in the absence of the particle. The bifurcation diagram is shown in fig.\ref{fig4}(a) and it is seen that upper part of the bifurcation curve of the soliton is dynamically stable with respect to linear perturbations. But with a particle placed in the center of the soliton the situation changes. We found stationary solution in the form of a resting soliton with a trapped particle placed in the center of soliton. The corresponding bifurcation curve is shown in fig.\ref{fig4}(b). The analysis of the dynamical stability reveals that the soliton is always unstable. In other words placing a particle in the center of the resting soliton destabilizes it. Such destabilization can be explained by the following: a particle creates a shadow and the soliton is located in the minimum of pumping intensity, but since the soliton is attracted by the region with higher pumping intensity the position of the soliton is no longer stable. It is important to note that the resting soliton destabilizes regardless of the transparency of the particle, what can be seen from \ref{fig1}(b), where the increment of linear excitations in the region of existence is shown.

The resting soliton with a captured particle in the center of soliton can be stabilized via using pumping field dependence on the coordinate in the form $P = P_0 e^{i k x^2}$. From the bifurcation diagram in fig.\ref{fig4}(c) it is can be seen that upper part of the bifurcation curve of solitons is partly stable. With increasing $k$ the upper soliton branch become fully stable. We found that in this case stability of the soliton depends on the transparency of a particle, or on the term $f$ in the equation (\ref{eq:schrod}). If particle shadow is too deep or in other words $f$ is larger some critical value $f_{0}$ then soliton collapses. Otherwise the soliton with a captured particle is stable and stays at rest. The dependence of the increment of linear excitations on the transparency of the particle can be seen in \ref{fig1}(c).

\begin{figure}[tbp]
\centering
\fbox{\includegraphics[width=\linewidth]{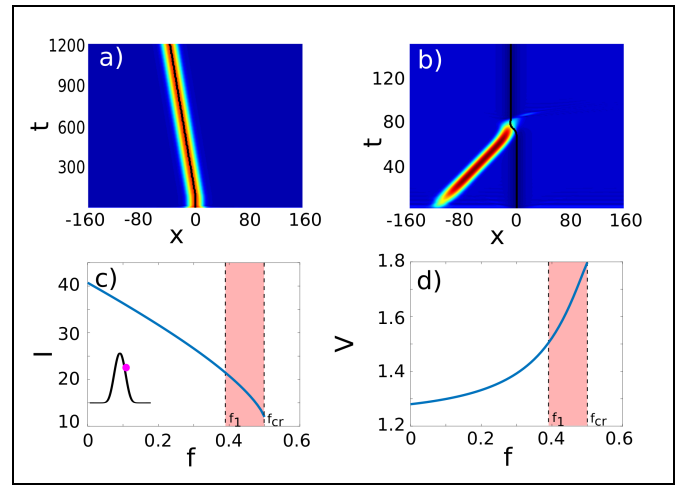}}
\caption{Fig.3 (a) The soliton starts to move because of the destabilization of the resting state by the particle, $P = P_0$, $f = 0.3$; (b) Formation of a moving soliton and further collapse due to the collision with the particle with $f = 0.5$; (c) Peak intensity of the moving soliton with the captured particle $I = |E|^2$; red area shows unstable solutions; pumping field has phase dependence on coordinate in the form $P = P_0 e^{i k x}$, where $k = 0.04$; (d) Dependence of velocity of the soliton with a captured particle on the transparency of the particle.}
\label{fig2}
\end{figure}

We studied the development of the instability of the resting soliton with a trapped particle under uniform pumping and we found out that the instability displaces the particle from the center of the soliton and sets the soliton in motion, see fig.\ref{fig2}(a). This motion caused by two gradient forces: first one attracts the particle to the soliton and second one repulses the soliton from the shadow of the particle. The moving solitons with captured particles will be considered in details in the next chapter.

\section{MOTION OF THE SOLITONS WITH TRAPPED PARTICLES}

Let us consider capturing a single particle by a moving soliton. The presence of the phase gradient of the holding beam sets dissipative solitons to motion. We choose the pump in the form of $P = P_0 e^{i k x}$, found numerically stationary solutions of moving solitons with captured particles and studied it's dynamical stability and changing the soliton velocity. It turned out that velocity and dynamical stability depend on the losses introduced by a particle, see \ref{fig2}(c-d). In the case of a moving soliton a particle is displaced from the center of the soliton and the bigger $f$ of the particle the bigger gradient of the pump intensity created by the particle shadow, that is why the velocity grows with increasing $f$. But there is a threshold value of $f = f_1$ at which the soliton gets destabilized. With further increasing $f$ to $f_{cr}$ stationary solution doesn't exist.

\begin{figure}[tbp]
\centering
\fbox{\includegraphics[width=\linewidth]{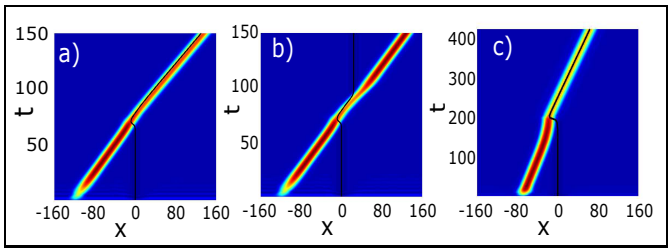}}
\caption{Fig.4 (a) Successful capture of a particle by a moving soliton, $f = 0.3$, other parameters are the same as in \ref{fig2}(c-d); (b) Capture of a particle by a soliton with further releasing the particle, $f = 0.4$; (c) Successful capture of a particle from (a) with a slower soliton $k = 0.01$;}
\label{fig3}
\end{figure}

We performed numerical simulations with the parameters insuring the existence of the soliton, which propagates with a constant speed and found out that depending on the stability of the corresponding stationaty states, when a soliton collides with a particle, several scenarios of further dynamics are possible. If the corresponding stationary solution is stable, soliton with a captured particle propagates together, otherwise instability leads to either loss of the particle by the soliton or to a collapse of the soliton. Collision of a soliton with a particle which introduces very large pumping losses leads to the destruction of the soliton, see \ref{fig2}(b). The case when the interaction of a soliton with a particle doesn't lead to destruction of the soliton is more interesting in terms of practical application. If the interaction with a particle doesn't lead to the soliton collapse, than crucial factor for further dynamics is the maximum value of the gradient force acting on the particle and the velocity of the soliton. If this force is large enough to ensure the motion of the particle with the velocity equal to the velocity of the soliton, then the corresponding stationary state is stable and the particle is captured by the soliton, see fig.\ref{fig3}(a). Otherwise, the soliton will pass by the particle causing its displacement, but not capturing it, see fig.\ref{fig3}(b). It can be explained by the fact that the capture of a particle leads to decreasing the pump intensity for a soliton, and although the force from the initial soliton is sufficient to capture the particle, trapping becomes impossible after particle' reducing the intensity of the soliton. In fig.\ref{fig3}(b) it can be seen that the capture 
of the particle occurs and for some time the soliton propagates with the particle, but, due to the fact that the particle reduces the pumping of soliton by creating a shadow, intensity of the soliton decreases, hence the gradient force acting on the particle decreases, and the particle gets released. But gradient force is able to ensure the motion of the particle with a slower velocity, so if the initial soliton moves slower, then capturing the same particle by the soliton is possible, see \ref{fig3}(c). The velocity of the soliton can be controlled by the phase gradient of a pump.

\section{CONCLUSION}

Now let us briefly summarize the main results of the paper. The dynamics of a particle under the action of an optical solitary wave in a pumped dissipative resonator with Kerr nonlinearity is considered.  Mathematically the dynamics of the system is described by Generalized Nonlinear Schrodinger equation for the optical field coupled to an ordinary differential equation describing the viscous motion of the particle. 

The stationary localized solution with trapped particles were found and their stability is analyzed. It was found that the particles can destabilize the solitons and depending on the parameters the instability can either destroy the soliton or to set the soliton in motion. It is also shown that the solitons with the trapped particles can be stabilized by spatially nonuniform holding beams.

The dynamics of the solitons colliding with the particles is investigated and it is shown that the collision can result in passing the particle through the soliton, in the trapping of the particle on the soliton or in the collapse of the soliton. 

The possibility to obtain and control the motion of the stable complexes consisting of solitons with trapped particles opens new possibility of optical manipulation of the nanoobjects driving a variety of emerging applications, such as, e.g., non-linear optical tweezers, nanoparticles delivery over arbitrary trajectories, optically induced governing chemical reactions in microchambers, etc.      

The work was supported by the Russian Foundation for Basic
Research (Projects No. 18-02-00414, 18-52-00005).
The calculations of the soliton dynamics are partially supported by the Russian
Science Foundation (Project No. 18-72-10127).

\end{document}